\documentclass[11pt]{article}
\usepackage{moriond,epsfig}

\def\ppbar{$p\overline{p} $}            
\def\bbbar{$b\overline{b} $}            
\def\qqbar{$q\overline{q}$}             
\def\nunubar{$\nu\overline{\nu}$}       
\def\met{\mbox{${\hbox{$E$\kern-0.6em\lower-.1ex\hbox{/}}}_T$}} 
\def\ipb{pb$^{-1}$}                     
\def\gev{GeV}                          
\def\gevc{GeV/c}                        
\def\gevcc{GeV/$c^2$}                   
\def\D0{D\O}                            
\def\et{\mbox{$E_T$}}

\def\stop1{$\widetilde t_1$}             
\def\gravitino{$\widetilde G$}             

\def\tauh{\mbox{$\tau_h$}}

\def\mtran{$M_T$}                        
\def\pt{$P_T$}                           
\def\htran{$H_T$}                       
 \def\mdim{$M_D$}                         
     
\def\sbar{$\overline{\widetilde t_1}$}   
\def\mstop{$m_{\widetilde t_1}$}                          
\newcommand{\rpv}{\mbox{$\not \hspace{-0.15cm} R$}}
\def\be{\begin{equation}}
\def\ee{\end{equation}}
\def\bea{\begin{eqnarray}}
\def\eea{\end{eqnarray}}

\begin{document}
\vspace*{4cm}
\title{Search for Hidden Dimensions in Run I Tevatron Data}

\author{ S. Hagopian }

\address{Department of Physics, Florida State University,\\
Tallahassee, Florida, 32312, U.S.A. \\
(on behalf of the \D0 and CDF collaborations)}

\maketitle\abstracts{
We describe the results of four new searches based on data samples from the
1992$-$1996 \ppbar   ~Collider run at Fermilab at $\sqrt{s}$ = 1.8 TeV. \D0 has searched 
for resonant slepton production in the two muon plus two jet channel. 
Using the   $R$-parity violating (\rpv) ~mSUGRA model,
{\ $\tilde \nu$ and $\tilde \mu$  }  masses up to 280 GeV/c$^2$
are excluded. CDF has searched for \rpv ~scalar top quark decays in two 
tau plus two jet events and set a limit of \mstop $ > 119 $ \gevcc.
\D0 has  searched for evidence of large  extra dimensions in monojet
production. Limits are set on \mdim, the fundamental mass scale, of 0.68  and
0.63  TeV for 
 $ n$ $=$ 4 and 6 dimensions. CDF has interpreted a search for new physics in 
 photon + missing energy events as a search for a gravitino and alternatively
 as a search for effects of graviton production. A gravitino mass limit $m_{3/2} > 1.17 \times 10^{-5}$ eV 
 is set. Limits on \mdim   ~of 0.55, 0.58, and 0.60 TeV
 ~for $ n$  $=$ 4, 6, and 8 are also determined.    
 }

\section{Introduction}

  The Standard Model (SM) of particles and forces provides a solid framework for 
  interpreting all the experimental data obtained so far. But it gives no
explanation for the great difference between the Planck scale of $10^{19}$ GeV
or the unification scale of $10^{16}$ GeV where all forces except gravity have equal 
strength, and the electroweak scale of 1 TeV where only the weak
and electromagnetic forces are unified. This
is refered to as the ``hierarchy problem.'' Supersymmetry (SUSY), which is a symmetry
between fermions and bosons, has been proposed as 
a solution to this problem. It predicts superpartners for all known particles.
The cancellation of terms between superparticles solves the Higgs mass
radiative correction problem.
$R$-parity, a  quantum number is defined which is $+1$ for SM particles
and $-1$ for SUSY particles \cite{rparity}. 
$ R = (-1)^ {3B+L+2S}$ where $B$,
$L$ and $S$ are the baryon, lepton and spin quantum numbers. 
 If $R$-parity is conserved then
SUSY particles are produced in pairs and decay to the lightest supersymmetric
particle (LSP).
But $R$-parity is not required by SUSY to be conserved. 

In supersymmetry, $R$-parity violation (\rpv) can occur through the 
following additional terms in the superpotential,
which are trilinear in the quark and lepton superfields \cite{rparity}:
\begin{eqnarray}
\lambda _{ijk} L_i L_j \bar{E^c_k}+
\lambda ' _{ijk} L_i Q_j \bar{D^c_k}+ 
\lambda '' _{ijk} \bar{U_i^c} \bar{D_j^c} \bar{D_k^c} ,
\label{super}
\end{eqnarray}
where $i,j,$ and $k$ are family indices; $L$ and $Q$ are the SU(2)-doublet lepton and
quark superfields; $E$, $U$, and $D$ are the singlet lepton, up and down quark
superfields respectively.

$R$-parity is approximately conserved. Experimental limits have been placed on 
the  $\lambda , \lambda ' $, ~and $\lambda '' $
coefficients \cite{Drein,referbound}.
  To conform to these limits, usually only one coefficient is assumed to be 
  non-zero. 
  $R$-parity violation can occur in the production of only one SUSY particle,
in the decay of the LSP to standard particles or in both production and decay.

   Theories with large extra dimensions eliminate the hierarchy problem by
bringing quantum gravity down to the electroweak scale. In a model developed
by Arkani-Hamed, Dimopoulos, and Dvali \cite{ADD}, $n$ extra spatial
dimensions are assumed to be compactified.  Standard Model   gauge
forces are confined to a 3+1 dimensional brane, while gravitons can propagate
in the whole space. The compactification of these dimensions gives
rise to Kaluza-Klein towers of states which appear as massive gravitons in the
3+1 brane.  If such
gravitons are directly produced and leave the 3 spatial dimensions, the 
experimental signature is the apparent  non-conservation of energy and momentum, such as in
events with a single photon or a single jet. The studies below use the 
phenomenological convention of Guidice, Rattazzi and Wells \cite{GRW}.
The compactified space is assumed  to be a torus with radius $ R$ and
fundamental mass scale  \mdim.  Newton's constant  $G_N^{-1} = M_{Pl}^2 $
~is replaced by  $G_N^{-1}$ = 8$\pi  R^n M_D^{2+n}$, making the effective Planck
scale the same as the electroweak scale for distances less that R. 
 Constraints from
astrophysical and cosmological considerations \cite{cosmos} are stronger than limits from 
gravitational measurements \cite{grav} and
rule out models of with $n < 4$ large extra dimensions. 

   The \D0 and CDF experiments took data from 1992$-$1996 at the Tevatron \ppbar 
   ~Collider at Fermilab at $\sqrt{s}$ = 1.8 TeV. This data has been extensively 
   studied
to look for hints of SUSY and its hidden sectors, and more recently to look
for effects predicted by theories with large extra dimensions. The results of 
four new searches are presented below.

\section{\D0 Search for Resonant Slepton Production }

   The two muon plus two jet final state has been studied by \D0 using 
   94  pb$^{-1}$of data. The \D0 detector is described in detail
   elsewhere \cite{detD0}. The results of this study
have been interpreted in terms of the $R$-parity violating mSUGRA model
with single smuon and smuon neutrino s-channel production, assuming dominant 
 $\lambda '_{211}$ 
coupling ($L_2 Q_1 \bar{D^c_1} $). The 
production of a slepton (smuon or smuon neutrino)
results in either a chargino or a neutralino
together with either a charged lepton or a neutrino.
It is assumed that the chargino decays into a neutralino, 
 which decays 
into two jets and a muon, if $\lambda '_{211}$ is dominant. 
The muon sneutrino decay into a muon and a
chargino and the smuon into a muon and a neutralino lead to at least
two muons and two jets in the final state. 
 The results of the analysis are expressed in terms of the
 5 mSUGRA parameters: $m_0$, the universal scalar mass at the unification scale
 $M_{GUT}$;
$m_{1/2}$, the universal gaugino mass at $M_{GUT}$;
$A=A_{t}=A_{b}=A_{\tau}$, the trilinear Yukawa coupling at $M_{GUT}$,which is
assumed to be zero,
sign$(\mu)$, the sign of the Higgsino mixing parameter and 
$\tan \beta$, the ratio of the vacuum expectation values of the Higgs fields.

\begin{figure*}
\begin{minipage}{0.45\linewidth}
\begin{center}
\mbox{\includegraphics[height=55mm] {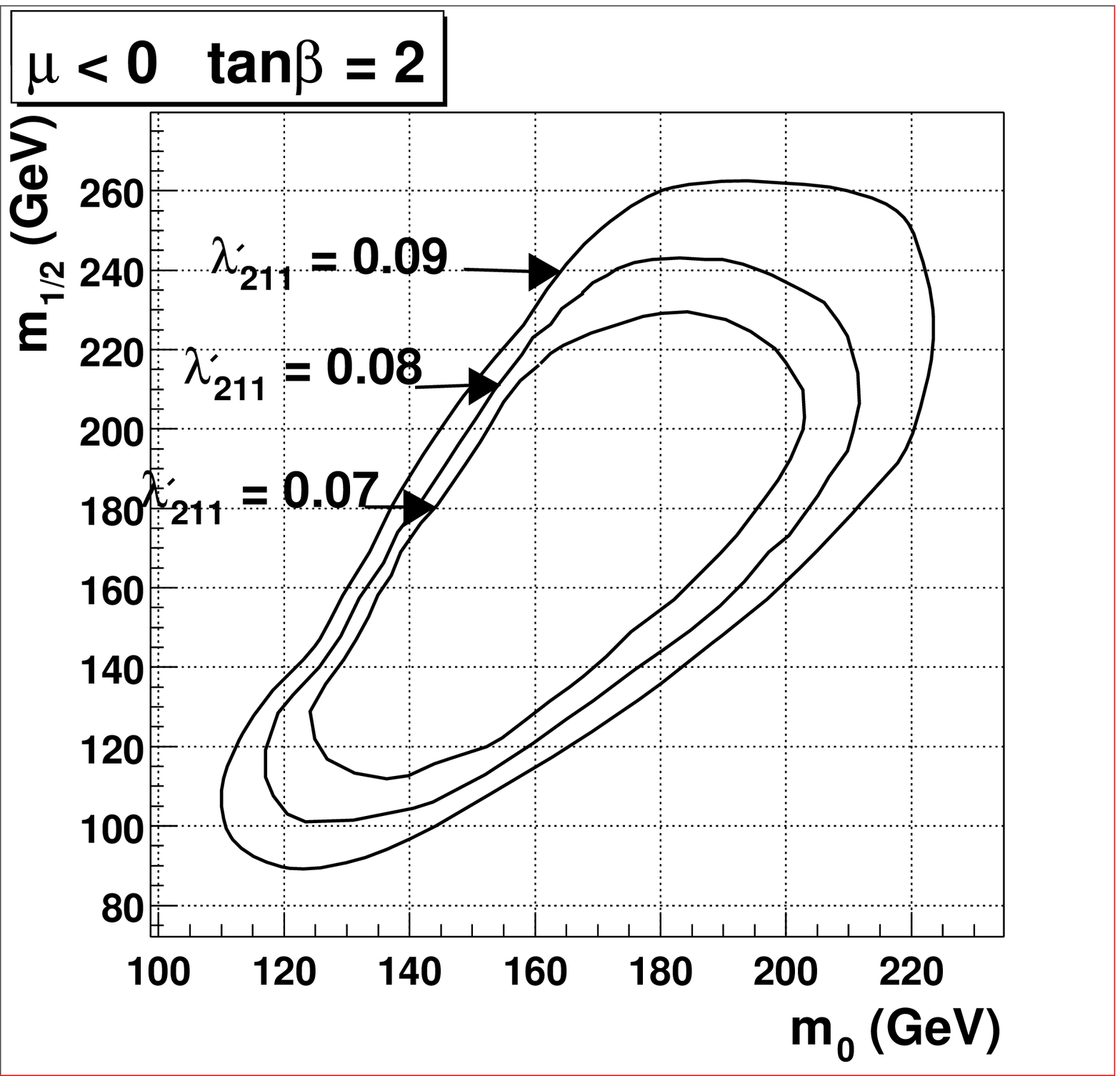}}
\end{center}
\end{minipage}
\hfill
\begin{minipage}{0.43\linewidth}
\begin{center}
\mbox{\includegraphics[height=55mm] {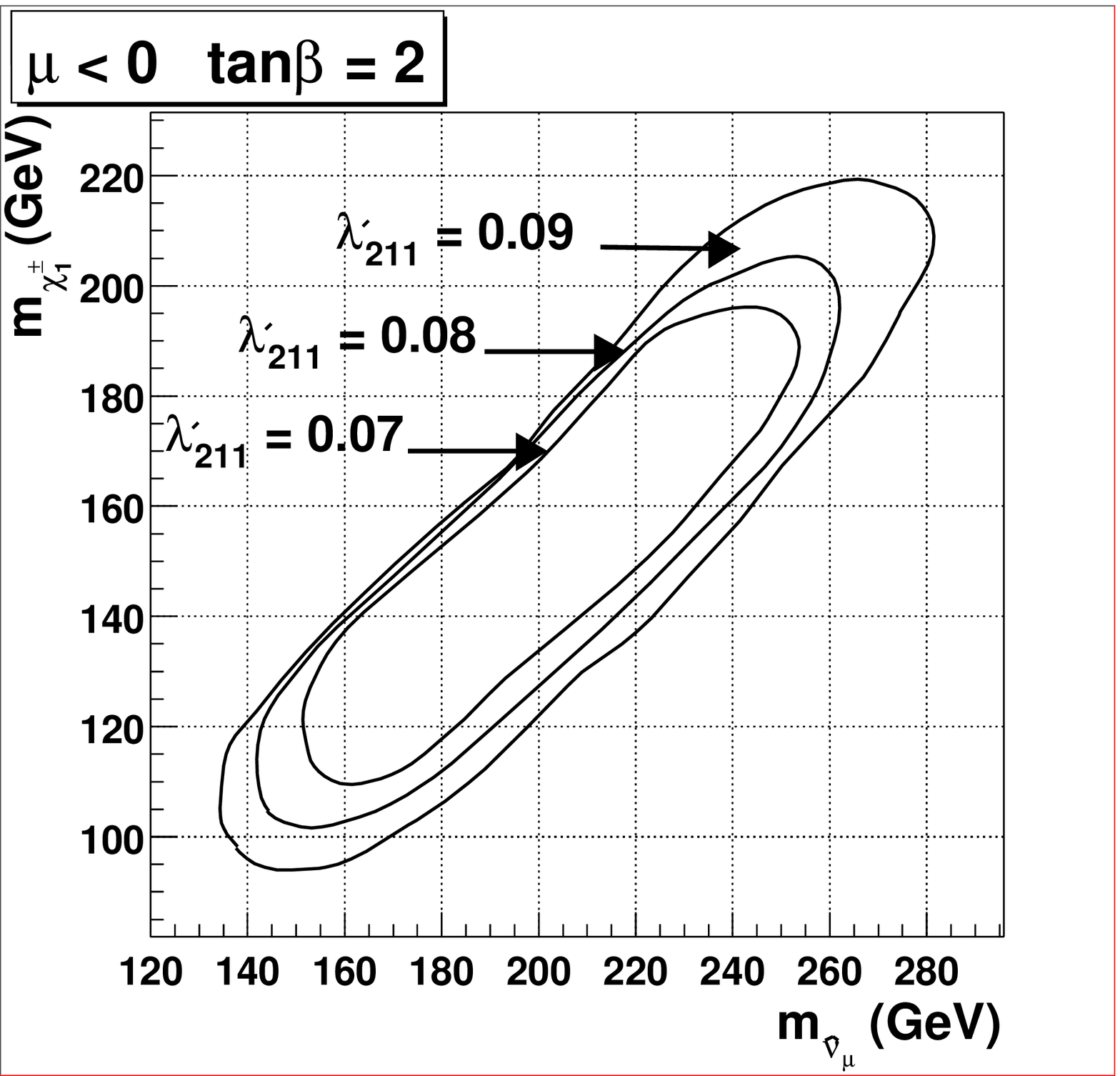}}
\end{center}
\end{minipage}
\hspace{7.0cm} (a) \hspace{7.0cm} (b)
\begin{flushleft}
\caption{(a) 
\D0 exclusion contours at the 95\% C.L. in the ($m_0$, $m_{1/2}$) 
plane for $\tan \beta =2$ and
$\lambda^{\prime}_{211}=$0.09, 
0.08 and 0.07. 
(b) Exclusion contours at the 95\% C.L. in the 
($m_{\tilde \mu}$/$m_{\tilde \nu}$, $m_{\chi^+}$) plane for
$\tan \beta=2$ and three values of $\lambda^{\prime}_{211}$.
Note that
all contour plots  are given as a function of the sneutrino mass. Due to the
fact that the sneutrino mass is very close to the smuon mass for a 
given set of parameters, these contours are very close to the sneutrino 
contour plot and thus are not plotted.
}
\end{flushleft}
\label{fig:fig1}      
\end{figure*}

The initial data selection consists of two central jets, with 
 $|\eta| < 1.0$ and $ \et > 20.0 $ ~GeV  and two central muons
 using   94 \ipb ~of data. The sum of the
 \et ~of the jets
is also required to be greater than 50 GeV.

\begin{figure}[tb]
\begin{center}
\epsfxsize=80mm
\epsffile{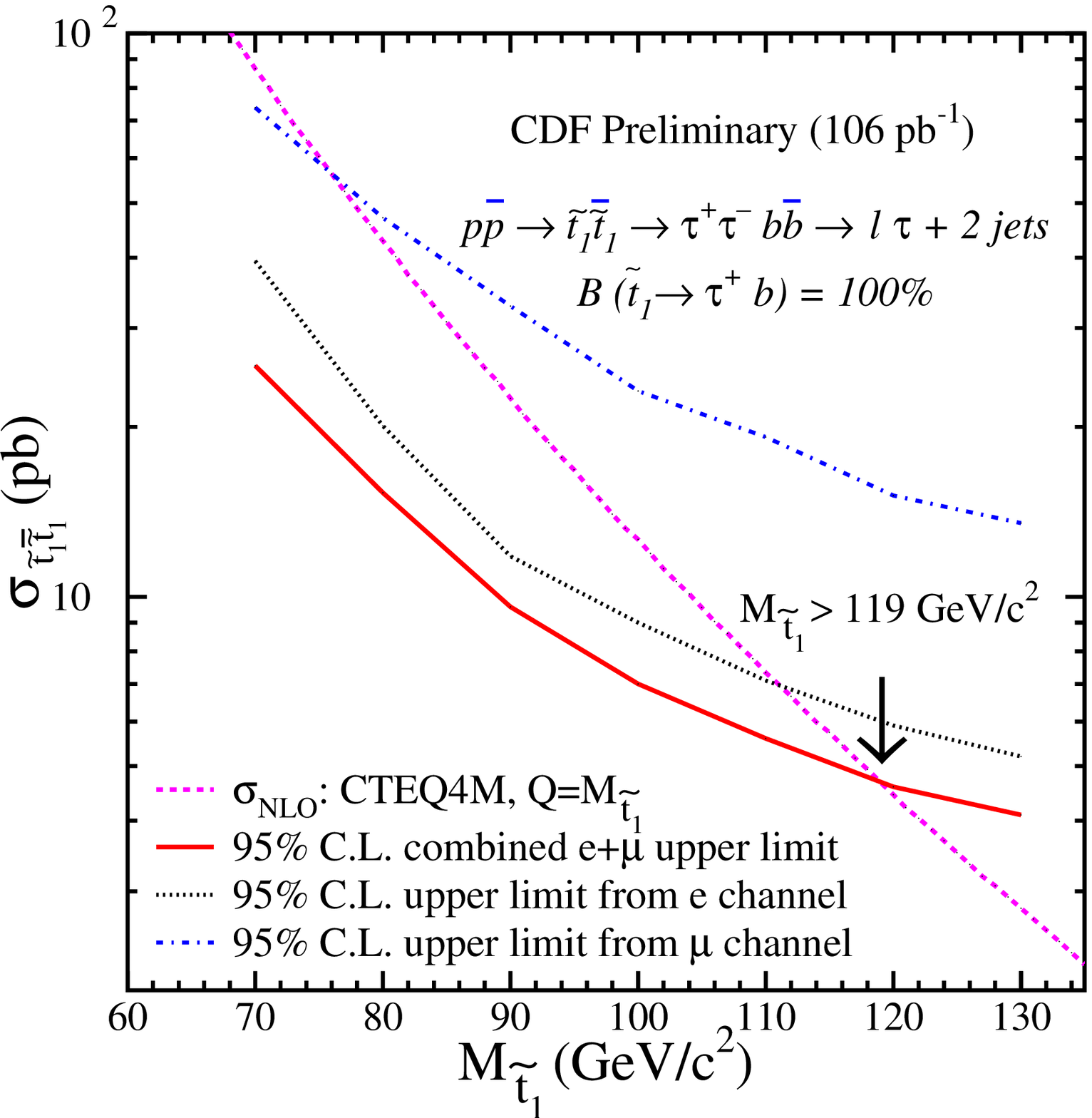} 
\vspace*{-0.05in}
\caption{
CDF 95\% C.L. upper limit  \stop1\sbar ~cross section from the e channel,  
    95\% C.L. upper limit  \stop1\sbar ~cross section from the $\mu$  channel,
   and 95\% C.L. combined e+$\mu$ upper limit cross section
for \stop1\sbar $\rightarrow \tau \tau $
\bbbar  ~and a NLO QCD cross section calculation using PROSPINO and CTEQ4M.
}
\vspace*{-0.25in}
\label{fig2}
\end{center}
\end{figure}
   
    The signal topologies were generated with the SUSYGEN Monte Carlo 
program~\cite{SUSYGEN} using the cross sections computed in two papers
by F. D\'eliot $et~ al.$
\cite{ourpap1,ourpap} for a wide range of  ($m_0$, $m_{1/2}$) 
masses. 
  The analysis is further refined using a 
 7--5--3--1 neural network with the input variables:
the sum of
the transverse energies of the two leading jets, the sum of
the transverse momenta of the two leading muons, 
the distance in ($\eta$, $\phi$) 
space between the two muons, the leading dimuon mass,
the ($\eta$, $\phi$) distance between the most energetic muon and its 
nearest jet, the sphericity in the laboratory frame, and 
the aplanarity. The neural network parameters  are trained for desired results 
of -1 for background and +1 for signal events.
Applying the best signal-over-background ratio to the neural network output
gives optimized cuts at $0.0$ for the $\tilde \nu$ and $-0.10$ for the 
$\tilde \mu$, respectively. The overall efficiency is 8\% for $\tilde \nu$
and 4\% for $\tilde \mu$.
The estimated background is $1.01 \pm 0.02$, mostly from $Z$+2 jets
   and $t\bar{t}$, with a small contribution from $WW$ + jets and is consistent
   with the 2 events observed.
 A Bayesian method is used to calculate the exclusion
confidence levels (C.L.). 
To set 
exclusion contours, the $(m_0$, $m_{1/2})$ plane is scanned for three 
values of the coupling constant 
$\lambda^{\prime}_{211} = 0.09,\ 0.08,\ 0.07 $, with
   $\tan \beta=2$, and   
$\rm{sign}(\mu) =-1$. The 
resulting exclusion contours at the 95\% C.L. 
are shown in the ($m_0$,\ $m_{1/2}$) plan in Figure 1 (a)  and    
in the ($m_{\tilde \mu}$/$m_{\tilde \nu}$, $m_{\chi^+}$) plane in Figure 1 (b). 
Values of  $m_{1/2}$  up to 260 {\rm GeV/c$^2$} for 
$\tan \beta$=2 and $\lambda^{\prime}_{211} = 0.09$ are excluded.
Equivalently masses up to 280 {\rm GeV/c$^2$
of $\tilde \nu$ and $\tilde \mu$  } are excluded.  
 
For low values of $m_0$ and $m_{1/2}$, the smuon mass is close to the chargino
or neutralino mass, the $p_T$ spectrum of the muons is soft, and the search
is inefficient. For $\mu>$0 and higher values of $\tan \beta$, the sensitivity 
decreases because
the photino component of the LSP becomes small, resulting in the decrease
of the branching fraction of the LSP into muons.
In addition, charginos and neutralinos become light, resulting in events
with softer muons and jets that fail the kinematic requirements.

\section{CDF Search for $R$-parity Violating Scalar Top Quark Decays in Two Tau
and Two Jet Events }

 A search for the lightest scalar top quark (\stop1), a supersymmetric
 partner of the top quark, has been performed using  the 105.3 \ipb ~data
 collected by the CDF detector, which is described elsewhere
 \cite{detCDF}. It is assumed 
 that the scalar top quarks are produced in pairs and each decays through
the $\lambda '_{333}$   \rpv ~coupling: \stop1 $\rightarrow \tau$ + b. This implies that
  $R$-parity violation happens only in the third generation 
($L_3 Q_3 \bar{D^c_3} $  superpotential term). 

    The experimental 
signature is two tau leptons and two b jets, where one tau decays into
either an electron or a muon and the other tau decays into hadrons.
The electron is required to have $ \et > 10$ GeV,
 ${p_T}^{trk} > 8$ ~\gevc, and  $|\eta| < 1.0$.
The muon is required to have ${p_T} > 10$~\gevc  ~and  $|\eta| < 0.6$.
The second tau, decaying into hadrons (\tauh), is required to have ${p_T} > 15$ ~\gevc 
~and  $|\eta| < 1.0$. The \tauh ~must have 1 or 3 tracks in a narrow
cone and meet the CDF tau identification cuts \cite{cdfwtau}.
Accepted events must also have  two or more jets with \et $> 15$ ~GeV. 
Additional selection criteria were imposed on transverse mass and 
scalar \et : \mtran ({\itshape l}, \met) $<$ 35 ~\gevcc ~and 
\htran ({\itshape l}, \tauh, \met) $>$ 75 ~\gev, where {\itshape l} = e or $\mu$.
After applying these cuts no event
is observed in either the electron or muon channels. 
The expected background is $1.92 \pm 0.18$ for the electron channel
and  $1.13 \pm 0.14$ for the muon channel from $Z(\rightarrow\tau\tau)$+jets,
   $W$+jets, diboson, Drell-Yan, top, and multijet events.
   The acceptance for \mstop =  120 ~\gevcc  ~is 3.18\% ~for electron data
   and 1.79\% for muon data. The cross section of ~\stop1\sbar ~events
   is calculated by normalizing to Z $\rightarrow \tau \tau $ events, which
   cancels some systematic uncertainties.
   A 95\% C.L. upper limit  \stop1\sbar ~cross section from the e  channel,  
   a 95\% C.L. upper limit  \stop1\sbar ~cross section from the $\mu$  channel,
   and a 95\% C.L. combined $e+\mu$  upper limit cross section
   are shown in Fig. 2 along with the next-to-leading order (NLO)
   QCD calculation using PROSPINO \cite{prospino} and CTEQ4M. This gives a limit of \mstop $ > 119 $ \gevcc.

\section{\D0 Search for Evidence of Large  Extra Dimensions in Monojet
Production} 

   A search has been made for evidence of real graviton emission  by \D0
using   78.8 \ipb ~of data. Events are required to have one jet with  $ \et >
150.0$ ~GeV and
\met $ > 150.0 $ ~GeV. Events are rejected if
there is a second jet with $ \et > 50.0$ ~GeV or if the event has an isolated muon.
The background is estimated to be  $38.0 \pm 9.6$, with $30.2 \pm 6.4$
from $WZ$ production and the remaining contributions from QCD and cosmics. This
is consistant with 38 events remaining in the data. 
  Graviton Monte Carlo signal events were generated using a subroutine provided by
 J. Lykken and K. Matchev \cite{match} as the external processs in
 PYTHIA \cite{PYTHIA}. 1000 events were generated for a range of $ n$ dimensions
 from 2 to 7, with the fundamental mass scale, \mdim, ~ranging from
 600 GeV to 1200 GeV, in 200 GeV intervals. The signal and background monte
 carlo samples were processed through the D0 Fast Monte Carlo Simulation
 Program.
The acceptance for the signal varies between 5\% ~and 8\% depending on
the values of $ n$ and \mdim.
Upper limits on
the cross section for the production of gravitons are calculated using
Bayesian statistics and assuming no signal contribution in the data. 
Limits assuming a K factor \cite{kfactor} of 1.34  are also calculated. From these cross
sections, lower limits on \mdim ~are determined for different interger values of $ n$.
These are shown in Fig. 3 and listed in Table 1. The \D0 limit
is less than the best LEP limit from DELPHI \cite{gDELPHI} of \mdim $>$ 0.84 TeV for $n =$ 4.
 But the higher center-of-mass
energy of the Tevatron leads to a better \D0 limit of 0.63 TeV for $n =$6, compared
to the DELPHI limit of 0.58 TeV.

\begin{figure}[tb]
\begin{center}
\epsfxsize=70mm
\epsffile{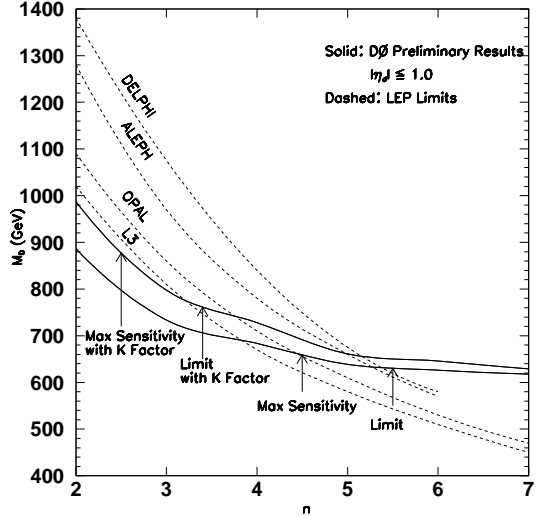} 
\vspace*{-0.05in}
\caption{
\D0 95 \% C.L. exclusion contour for large extra dimensions with leading
jet  $ \et > 150.0$ ~GeV, \met $> 150.0$ ~GeV, and $|\eta| \leq 1.0$.
Dashed curves are the LEP limits .
}
\vspace*{-0.050in}
\label{fig3}
\end{center}
\end{figure}

\begin{center}
\begin{table}[htb]
\vspace*{-0.1in}
\caption{Lower limits at 95\% C.L. on the fundamental mass scale, \mdim, in
TeV.} 
\vspace*{+0.1in}
\label{table1}
\begin{center}
\begin{tabular}{c|@{}cccccc}
\hline
   no. of extra dimensions & ~~$n$=2 & $n$=3 & $n$=4 & $n$=5 & $n$=6 & $n$=7~~  \\
\hline
\mdim (TeV) & ~~0.89 & 0.73	& 0.68	  & 0.64   & 0.63     & 0.62~~  \\
\hline
\end{tabular}
\end{center}
\vspace*{-0.3in}
\end{table}
\end{center}

\section{CDF Limits on Extra Dimensions and New Particle Production in Photon +
Missing Energy Events}
 
   Single photon production can indicate the production of invisible
 particles such as the gravitino in SUSY models and the graviton in
 models with large extra dimensions. 
 CDF has studied the  photon + \met ~channel  using   
87  \ipb ~of data. The main selection criteria are one electromagnetic (EM) jet 
with \et$^\gamma > 55.0$  \gev ~and $|\eta^\gamma | < 1.0$. The EM
jet must pass CDF photon identification cuts \cite{gCDF,cdfgamma}. Additional 
requirements
include \met $>$ 45 GeV, no jets with \et $>$ 15 GeV and no tracks with \pt $>$ 5 GeV.
Special timing cuts were also made to reduce cosmic ray events.
The overall efficiency for all cuts varies from 45\% at \et $=$ 55 GeV to 56\%
for \et $>$ 100 GeV.

The SM background is small for this signature. The largest background,
$6.2 \pm 2.0$ events, is due to remaining cosmic ray muons that
bremsstrahlung a $\gamma$ in the EM calorimeter, but no corresponding track
is found. The second largest background from \qqbar $\rightarrow Z\gamma \rightarrow $ \nunubar $\gamma$
 is $3.2 \pm 1.0$
events. Small background contributions also come from W $\rightarrow e\nu$,
prompt diphoton production, and $W\gamma$ production. The QCD background
is estimated to be less that 1.0 events, but is not used since the
uncertainty is very large. This gives a more conservative limit. 
The total background of $11.0 \pm 2.2$ is consistant with the 11 events which remain after the cuts.

 Both background and signal events were generated using a modified version of PYTHIA with CTEQ5L  
and then processed through the CDF detector simulation program. 
   The lack of excess events can be interpreted as a limit on superlight
 gravitino production in the process \qqbar $\rightarrow$ \gravitino \gravitino
 $\gamma$, using the model of Brignole $et~ al.$ \cite{brignole}, 
which assumes that the gravitino is the LSP and that the other supersymmetric
particles are too massive to be produced at Tevatron energies. 
This sets an
absolute lower limit on the gravitino mass, $m_{3/2}$ and supersymmetry breaking
scale  $|F|^{1/2}$, where
$|F| = \sqrt{3}
m_{3/2}M_{Pl}$.  From the lack of excess events in 
$\gamma +$ \met,
a  95\% C.L. limit of $|F|^{1/2} >$ 221 GeV is set. The is equivalent to a gravitino
mass limit $m_{3/2} > 1.17 \times 10^{-5}$ eV, which is similar to 
previous limits set  by DELPHI in
a $\gamma +$ \met  ~study  and by CDF in a monojet $+$ \met ~analysis \cite{gDELPHI,gCDF}.

   The same channel can also be interpreted as a search for direct graviton
production from quark-antiquark annihilation. The limits obtained
are listed in Table 2.
\begin{center}
\begin{table}[htb]
\vspace*{-0.1in}
\caption{Lower limits at 95\% C.L. on the fundamental mass scale, \mdim, in TeV.}
\vspace*{+0.1in}
\label{table2}
\begin{center}
\begin{tabular}{c|@{}ccc}
\hline
   no. of extra dimensions & ~~$n$=4 & $n$=6 & $n$=8~~  \\
\hline
\mdim (TeV) & ~~0.55 & 0.58     & 0.60~~  \\
\hline
\end{tabular}
\end{center}
\vspace*{-0.3in}
\end{table}
\end{center}

\section{Conclusions}
 The new searches of Tevatron data show no evidence of the effects of
 supersymmetry or large extra dimensions. Other Tevatron  analyses are now in progress.
 There is more good physics results to come from Run I data, while we look forward
 to Run II results!

\end{document}